\documentclass[prl,superscriptaddress,showpacs,twocolumn]{revtex4}
\usepackage{graphicx}

\def\be{\begin{eqnarray}}
\def\ee{\end{eqnarray}}
\newcommand{\ice}[1]{\relax}

\newcommand\VV{\setbox0=\hbox{V}\hbox{\rm V\raise\ht0
  \hbox to0pt{\hss\vbox to0pt{\hbox{v}\vss}}}}
\def\slashchar#1{\setbox0=\hbox{$#1$}           
   \dimen0=\wd0                                 
   \setbox1=\hbox{/} \dimen1=\wd1               
   \ifdim\dimen0>\dimen1                        
      \rlap{\hbox to \dimen0{\hfil/\hfil}}      
      #1                                        
   \else                                        
      \rlap{\hbox to \dimen1{\hfil$#1$\hfil}}   
      /                                         
   \fi}                                         %

\begin{document}

\title{Heavy mass expansion, light-by-light scattering and the anomalous
magnetic moment of the muon}

\author{J.H.~K\"uhn}

\affiliation{Institut f\"ur Theoretische Teilchenphysik,
Universit\"at Karlsruhe, D-76128 Karlsruhe, Germany}

\author{A.I.~Onishchenko}

\affiliation{Institut f\"ur Theoretische Teilchenphysik,
Universit\"at Karlsruhe, D-76128 Karlsruhe, Germany}

\affiliation{Institute for High Energy Physics, 142284
Protvino, Moscow region, Russia}

\affiliation{Department of Physics and Astronomy,
Wayne State University, Detroit, MI 48201, USA}

\author{A.A.~Pivovarov}

\affiliation{Institut f\"ur Physik, Johannes-Gutenberg-Universit\"at,
Staudinger Weg 7, D-55099 Mainz, Germany}

\affiliation{Institute for Nuclear Research of the Russian
Academy of Sciences, 117312 Moscow, Russia}

\author{O.L.~Veretin}

\affiliation{Institut f\"ur Theoretische Teilchenphysik,
Universit\"at Karlsruhe, D-76128 Karlsruhe, Germany}

\begin{abstract}

Contributions from light-by-light scattering to $(g_\mu-2)/2$, the
anomalous magnetic moment of the muon, 
are mediated by the exchange of charged fermions or 
scalar bosons.
Assuming large masses $M$ for the virtual particles and employing the
technique of large mass expansion, analytical results are obtained for
virtual fermions and scalars
in the form of a series in $(m_\mu /M)^2$.
This series is well convergent even for the case $M=m_\mu$.
For virtual fermions, the expansion confirms published
analytical formulae. For virtual scalars, the result can be used to
evaluate the contribution from charged pions.
In this case our result confirms already
available numerical evaluations, however, it is significantly more
precise.

\end{abstract}


\maketitle

High precision measurements of the muon anomalous magnetic
moment, $a_\mu=(g_\mu-2)/2$, are used for stringent quantitative
tests of the theories suggested for describing particle interactions.
Presently the world average
of the muon anomalous magnetic moment has a relative precision
of 0.7 parts per million (ppm)~\cite{Bennett:2002jb,previous}
\be
\label{expnumber}
a_\mu({\rm exp}) = 11 659 203(8)\times 10^{-10}\, .
\ee
and a further reduction of the experimental error by a factor two is
within reach.

In the standard model (SM) the theoretical value of $a_\mu$
is given by a sum of three contributions: $a_\mu({\rm SM}) =
a_\mu({\rm QED})+a_\mu({\rm weak})+a_\mu({\rm had})$.
The contribution from QED, $a_\mu({\rm QED})$, which includes those from
virtual leptons, and the one from weak interactions, $a_\mu({\rm weak})$,
can be uniquely evaluated in perturbation theory with the results:
$a_\mu({\rm QED}) = 11 658 470.57(0.29)\times 10^{-10}$~\cite{qedcont}
and $a_\mu({\rm weak}) = 15.1(0.4)\times 10^{-10}$~\cite{weak}.
The hadronic 
piece $a_\mu({\rm had})$, however,
is sensitive to long
distance effects and cannot be evaluated in  a perturbative framework.
Using the experimental value from Eq.~(\ref{expnumber})
one expects for the remaining hadronic contribution
$a_\mu({\rm had})=a_\mu({\rm exp})-a_\mu({\rm QED})
-a_\mu({\rm weak})=717(8)\times 10^{-10}$.
To test the standard model and to search for ``new physics'', this
value must be reproduced by a precise evaluation of the effects to be
discussed in the following.

The dominant contributions to $a_\mu (\rm had)$ are those from the hadronic
vacuum polarisation in lowest order (one particle irreducible part)
with most recent results
$a_\mu({\rm had},{\rm LO})=(702\pm6\pm14)\times 10^{-10}$~\cite{had1b},
$a_\mu({\rm had},{\rm LO})=692(6)\times 10^{-10}$~\cite{had1a},
$a_\mu({\rm had},{\rm LO})=683.6(8.6)\times
10^{-10}$~\cite{Jegerlehner:2001wq},
$a_\mu({\rm had},{\rm LO})=(683\pm5.9\pm2.0)
\times10^{-10}$~\cite{Hagiwara:2002ma}
and the range between
$a_\mu({\rm had},{\rm LO})=(684.7\pm6.0_{\rm exp}\pm 3.6_{\rm
rad})\times 10^{-10}$
and
$a_\mu({\rm had},{\rm LO})=(709.0\pm 5.1_{\rm exp}\pm 1.2_{\rm rad}\pm
2.8_{\rm SU(2)})\times 10^{-10}$ for analyses
based on $e^+e^-$ and $\tau$ data sets respectively~\cite{Davier:2002dy}.

The next-to-leading order receives one contribution from the
reiteration of the hadronic vacuum polarisation which is known
to be negative and can be calculated unambiguously and
with sufficient precision,
$a_\mu({\rm had,NLO;pol})=-10.1(0.6)\times 10^{-10}$~\cite{krause}.
Its sign  is strictly fixed:  the  kernel
is negative definite~\cite{barbieri} and the spectral density of
the
two-point correlator of the hadronic electromagnetic current is positive.
Another NLO contribution originates from  light-by-light scattering
which cannot be calculated on the basis of data or 
first principles.
The evaluations of this term have changed dramatically during last
years~\cite{Hayakawa:1996ki,Bijnens:1996xf,weak,change,nowlbl}, with
$a_\mu({\rm had,NLO;lbl})=8.6(3.2)\times 10^{-10}$
as the present conservative estimate based on the evaluation of hadronic
formfactors and resonance contributions~\cite{Bennett:2002jb,nowlbl}.
It is remarkable that
the approach of ref.~\cite{change} based on the concept
of quark-hadron duality with an effective quark mass accounting for
the QCD long distance effects leads to
$a_\mu({\rm had,NLO,lbl})_{\rm duality}=14(3)\times 10^{-10}$
in good agreement with the evaluations based on models for the
hadronic formfactors.

One important part of light-by-light-scattering amplitudes originates from
neutral, low mass intermediate resonances, dominantly $\pi^0$ and, less
important, $\eta$, another from charged pion loops. This second term
is the main subject of the present paper.

In view of the low mass of the muon and pion, as compared to the characteristic
scale of the pion form factor, $m_\rho^2 \approx 0.6~{\rm GeV^2}$, pions can be
treated as pointlike.
The corresponding problem of loops of (pointlike) fermions has been
solved in analytical form, with the final result expressed in the form of a
complicated expression of polylogarithms depending on $(m_f^2 /m_\mu^2)$
\cite{Laporta:1992pa}.

\begin{figure}[ht]
\quad\includegraphics[scale=0.3]{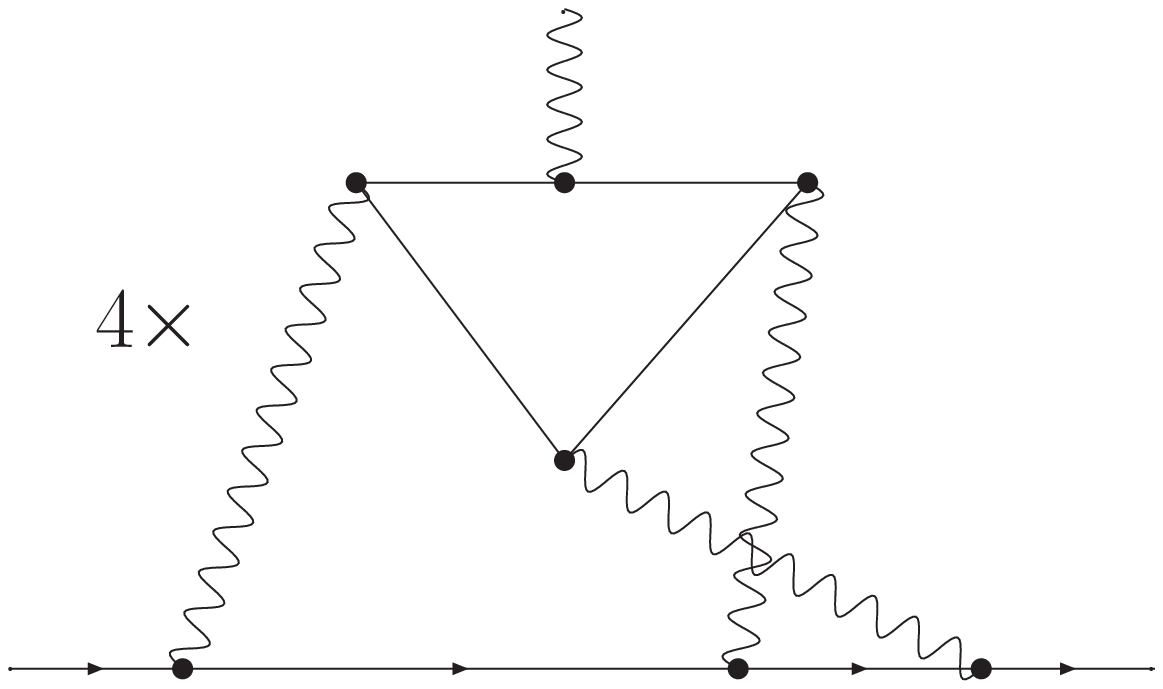} \makebox[1mm]{}
\includegraphics[scale=0.3]{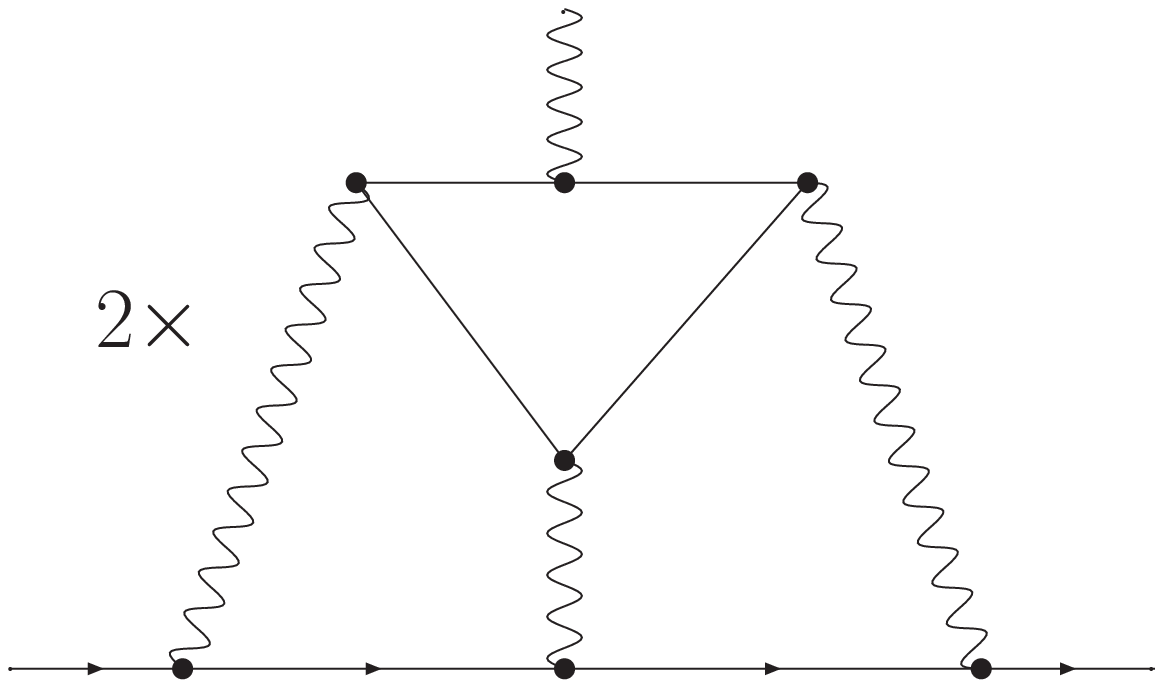}
\caption{QED-type light-by-light diagrams}
\label{QEDfig}
\end{figure}

In the present paper, a different approach is adopted, which makes use of
the fact that $m_\mu^2 /m_\pi^2 \approx 0.6$ is significantly smaller than 1.
Using the heavy mass expansion~\cite{asymptotic},
the result is obtained in the form of a
power series, which
can be systematically constructed up to arbitrary
orders in $(m_\mu/m_\pi)^2$. The corresponding  expansion can also be
constructed for heavy fermions, leptons as well as quarks.
The fermionic contribution to light-by-light
scattering is given by the 6 diagrams depicted in
Fig.~\ref{QEDfig} (with permutations of legs omitted). The analytical value
for this contribution is known in QED
for arbitrary values of $m$ and $M$
since long~\cite{Laporta:1992pa}. 
\vspace*{0.2cm}
Also the two leading terms in the
$m^2/M^2$ expansion can be found in this reference.
Below we reproduce these terms and give
three more terms of the expansion. To evaluate
the  contribution of charged pion loops we consider the pion as
an elementary pointlike particle and use the Lagrangian
\begin{equation}
L_{\rm sQED}=|D_\mu \pi|^2 - m_\pi^2 |\pi|^2 \ ,
\quad D_\mu = \partial_\mu-i e A_\mu
\end{equation}
to describe its interaction with photons. This is well justified
since the integral is convergent in the high energy
region, and the internal structure of the pion is not yet resolved for
momenta of order $m_\mu$ or $m_\pi$.

\begin{figure}[ht]
\begin{center}
\includegraphics[scale=0.2]{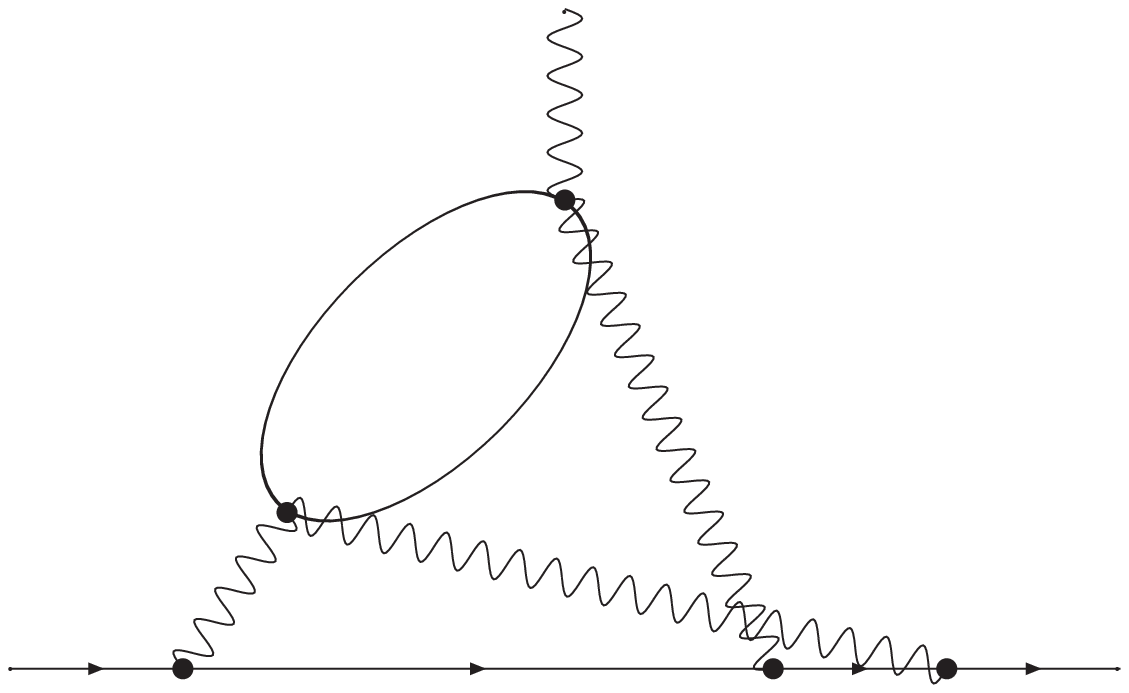} \makebox[4mm]{}
\includegraphics[scale=0.2]{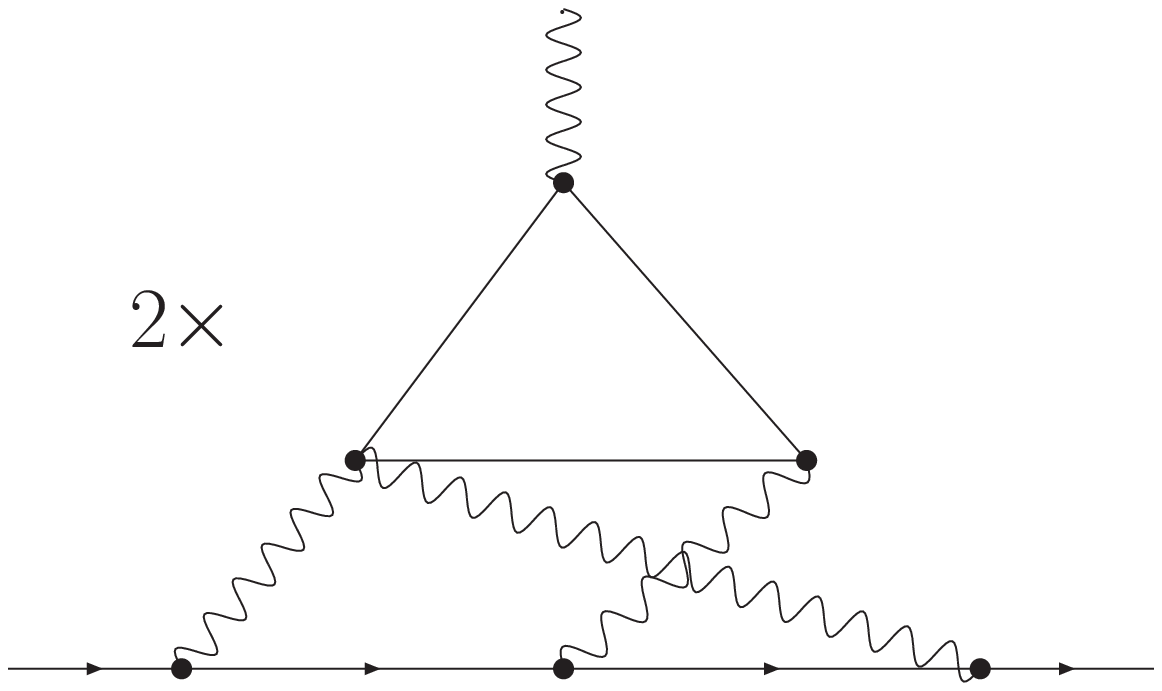}
\end{center}
\begin{center}
\includegraphics[scale=0.2]{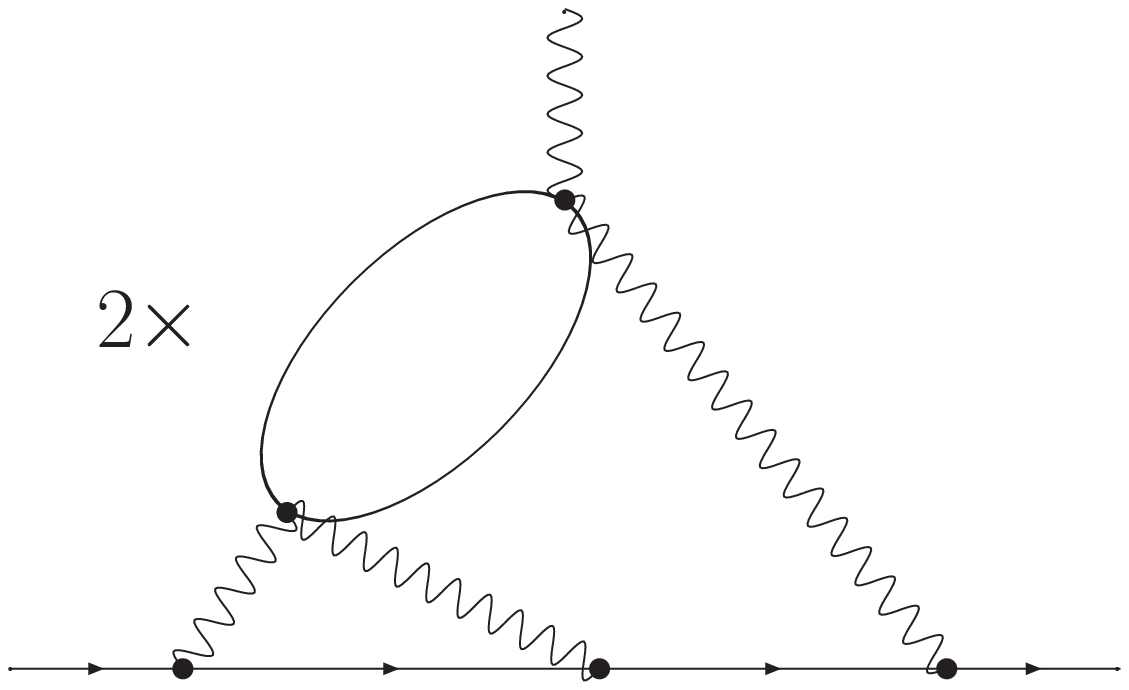} \makebox[2mm]{}
\includegraphics[scale=0.2]{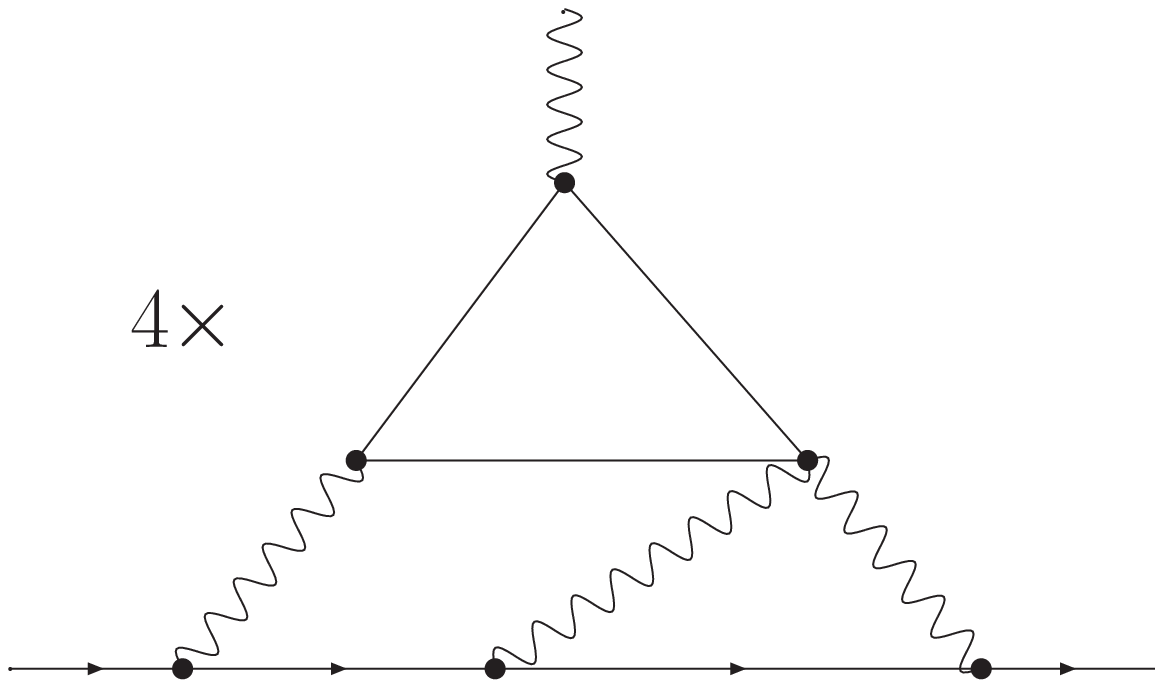}
\end{center}
\begin{center}
\includegraphics[scale=0.2]{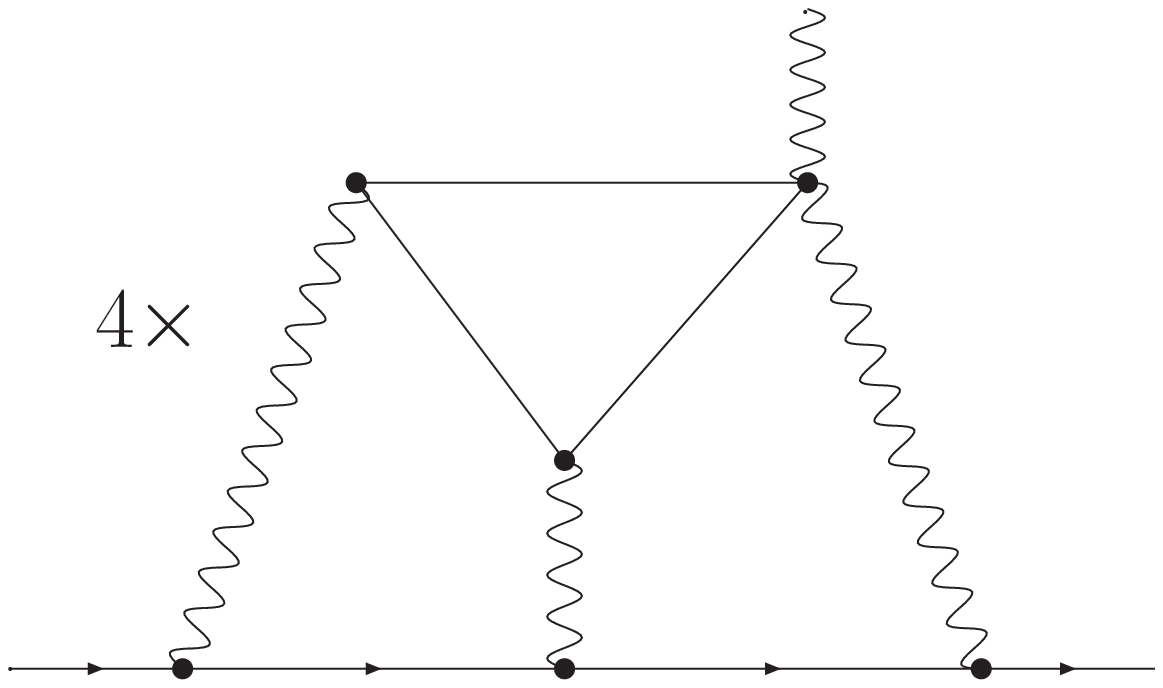} \makebox[5mm]{}
\includegraphics[scale=0.2]{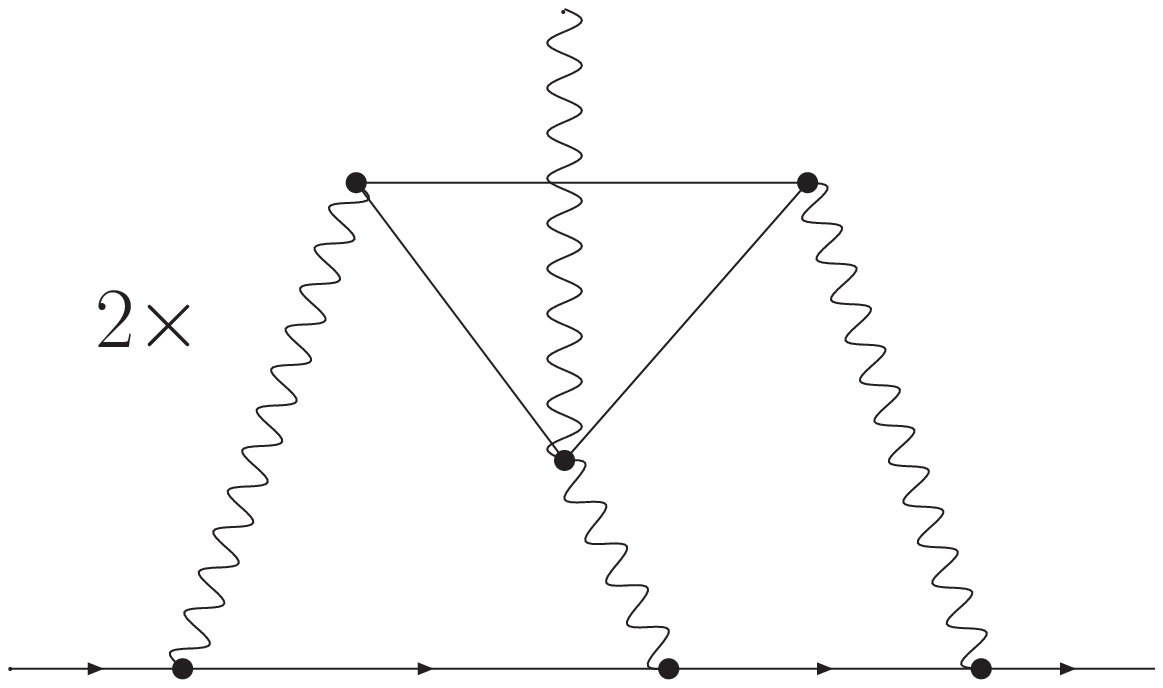}
\end{center}
\begin{center}
\quad\includegraphics[scale=0.2]{feyn7.eps} \makebox[1mm]{}
\includegraphics[scale=0.2]{feyn8.eps}
\end{center}
\caption{Scalar QED light-by-light diagrams}
\label{sQEDfig}
\end{figure}

The  21 diagrams for the light-by-light
contribution as derived from scalar QED
are displayed in Fig.~\ref{sQEDfig} (permutations of legs omitted). 
Numerical
results for this contribution were already reported
in refs.~\cite{Kinoshita:1984it,Hayakawa:1996ki}.
Analytical expressions, however, are presented in this work for the first time.
In view of the fact that this contribution in indeed comparable with the
expected experimental error, an independent confirmation seems
furthermore, highly desirable.

\begin{figure}[h]
\includegraphics[scale=0.4]{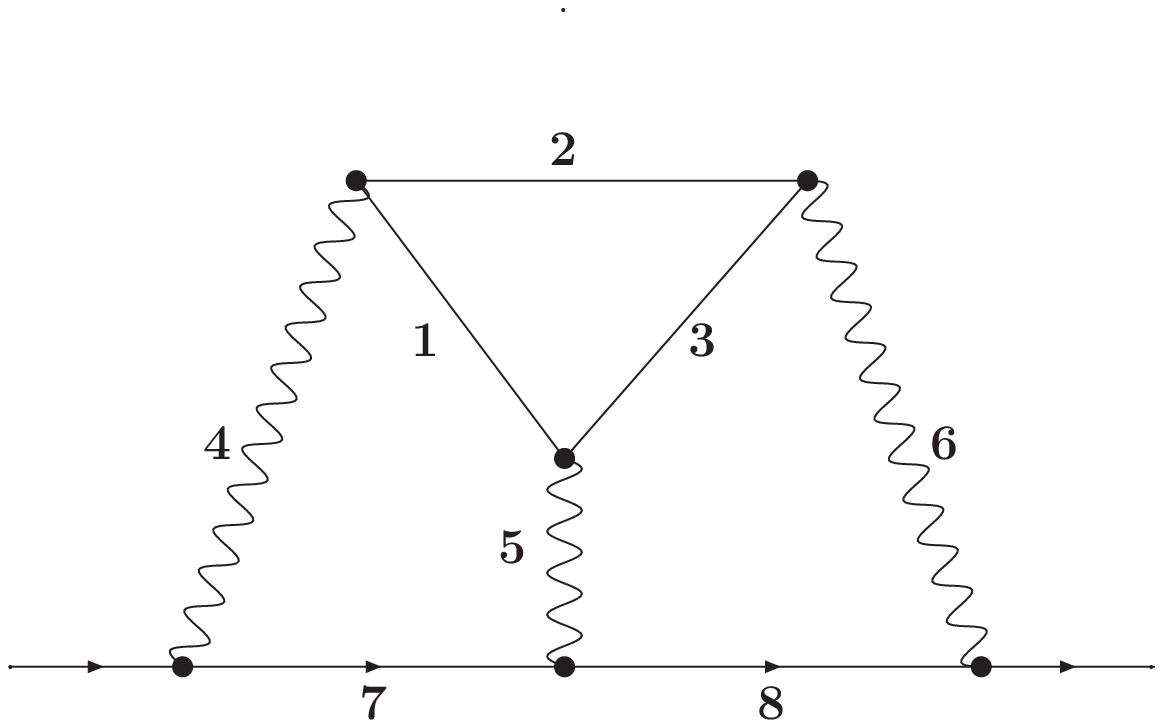}
\caption{A prototype diagram.
Lines 1, 2 and 3 have the mass of pion, lines 7 and 8 --- the
mass of muon while lines 4, 5 and 6 are masseless.
}
\label{prototype}
\end{figure}

To compute the contributions of different
diagrams we first project onto the relevant form factor of the anomalous
magnetic moment~\cite{proj}. Subsequently  we employ the well-known
method of asymptotic expansion
in the small mass ratio $m/M$~\cite{asymptotic}.
As an example let us 
consider the large mass expansion of the generic
diagram depicted in Fig.~\ref{prototype}. 
To obtain the
asymptotic expansion for a given diagram one has to compute
the sum of different contributions which are simpler than the original
Feynman integral. These different contributions can be classified
according to the so called "hard" subgraphs of the original diagram $\Gamma$.
These are defined as subgraphs containing all heavy lines such
that the corresponding co-subgraph remains one-particle irreducible.
The last step of
the heavy mass expansion consists of the  simple Taylor expansion of
the "hard" subgraphs in small masses and external momenta.
The generic prototype diagram without the external photon line
shown in Fig.~\ref{prototype}
leads to four different "hard" subgraphs:
1. the graph $\Gamma$ itself; 2. the subgraph formed by the lines
(1,2,3); 3. the subgraph formed by the lines (1,2,3,4,5,7);
4. the subgraph formed by the lines (1,2,3,5,6,8).
We now discuss these pieces separately.

1. In this case we perform a Taylor expansion of the diagram in
Fig.~\ref{prototype} in the external muon
momentum $p$ and the mass $m$. The resulting tensors
$p_{\mu_1}p_{\mu_2}\ldots p_{\mu_n}$ can be treated with the use
of the combinatoric formula
\be p_{\mu_1}p_{\mu_2}\ldots
p_{\mu_{2n}}&\to & \frac{\Gamma (d/2)}{2^n\Gamma (d/2+n)}(p^2)^n
g_{s,(n)} \nonumber \\
p_{\mu_1}p_{\mu_2}\ldots p_{\mu_{2n+1}}&\to& 0
\label{aver4}
\ee
where $g_{s,(n)}$ is the totally symmetric
product of $n$ metric tensors $g_{\mu\nu}$, $d$ is the space time
dimension within dimension regularization.
The resulting scalar one-scale 3-loop bubbles where computed using
the package MATAD~\cite{Steinhauser:2000ry}.

2. Here we perform the Taylor expansion in the momenta external to the
subgraph formed by lines (1,2,3). The resulting symmetric tensors built
from the loop momentum of the ``hard'' one-loop tadpole integrals are
expressed by the squared ``hard'' momentum with the help of
Eq.~(\ref{aver4}). The remaining scalar one-loop tadpoles are
expressed in terms of $\Gamma$-functions. Two-loop
scalar on-shell integrals are evaluated with the package
ON-SHELL2~\cite{onshell2}.

3. For this subgraph and similarly for subgraph (4)
we perform a Taylor expansion in the external momenta and
the muon mass $m$. The resulting tensor one-loop on-shell
integrals over soft momenta can be reduced to scalar integrals for which
simple explicit 
results exist (e.g.~\cite{Davydychev:1991va}).
Another way to handle the tensor structures that has been used for
a check of the calculation is the reduction of the tensor integrals for
a hard subgraph (explicit formulae are given in~\cite{Chetyrkin:rv}).
The remaining one-loop on-shell and two-loop single scale tadpole
integrals can be finally reduced to $\Gamma$-functions.

The steps described above were automized with the
Feynman diagram analyser DIANA~\cite{Tentyukov:1999is}
and the computer algebra system FORM~\cite{Vermaseren:2000nd}.

To discuss our results we introduce the standard normalization
for the light-by-light contribution
\be
a_\mu({\rm lbl}) = \left(\frac{\alpha}{\pi}\right)^3
a_\mu(\gamma\gamma)\, .
\ee
For the fermionic light-by-light contribution we find
\be
\label{qedres1}
&&a_{\mu}(\gamma\gamma;{\rm QED}) =
\frac{m^2}{M^2}\left( \frac{3}{2}\zeta_3-\frac{19}{16}\right)
\nonumber \\
&&
\hspace{-0.5cm}+\frac{m^4}{M^4}\left(\frac{13}{18}\zeta_3
-\frac{161}{1620}\zeta_2-\frac{831931}{972000}
-\frac{161}{3240}L^2-\frac{16189}{97200}L\right)\nonumber \\
&&+\frac{m^6}{M^6}\left(\frac{17}{36}\zeta_3-\frac{13}{224}\zeta_2
-\frac{1840256147}{3556224000} \right. \nonumber \\
&&\left.-\frac{4381}{120960}L^2-\frac{24761}{317520}L
\right)\nonumber \\
&& +\frac{m^8}{M^8}\left(\frac{7}{20}\zeta_3 -
\frac{2047}{54000}\zeta_2 - \frac{453410778211}{1200225600000}
\right. \nonumber \\
&&\left. -\frac{5207}{189000}L^2-\frac{41940853}{952560000}L
\right) \nonumber \\
&& +\frac{m^{10}}{M^{10}}\left(\frac{5}{18}\zeta_3 -
\frac{1187}{44550}\zeta_2- \frac{86251554753071}{287550049248000}
\right.\nonumber \\
&&\left. - \frac{328337}{14968800}L^2
-\frac{640572781}{23051952000}L \right)
+O\left(\frac{m^{12}}{M^{12}}\right)
\ee
where $L = \ln(M^2/m^2)$, 
$m$ and $M$ denoting muon and fermion mass respectively,
$\zeta_2=\zeta(2)=\pi^2/6$, $\zeta_3=\zeta(3)$.
The first two terms of this expansion coincide with the result given
explicitly in ref.~\cite{Laporta:1992pa}, the other terms are
new~\cite{thanksL}.

The expansion in scalar QED, relevant for the charged pion contribution,
has the form
\be \label{scalqedres1}
&&
a_{\mu}(\gamma\gamma;{\rm sQED})
= \frac{m^2}{M^2}\left(\frac{1}{4}\zeta_3
-\frac{37}{96} \right) \nonumber \\
&&\hspace{-0.8cm} +\frac{m^4}{M^4}\left(\frac{1}{8}\zeta_3
+\frac{67}{6480}\zeta_2
-\frac{282319}{1944000}+\frac{67}{12960}L^2
+\frac{7553}{388800}L\right) \nonumber \\
&&+\frac{m^6}{M^6}\left(\frac{19}{216}\zeta_3+\frac{157}{36288}\zeta_2
-\frac{767572853}{7112448000}\right. \nonumber \\
&&\left. +\frac{1943}{725760}L^2+\frac{51103}{7620480}L
\right)\nonumber \\
&&+\frac{m^8}{M^8}\left(\frac{11}{160}\zeta_3+\frac{943}{432000}\zeta_2
-\frac{3172827071}{37507050000} \right. \nonumber \\
&&\left.
+\frac{8957}{6048000}L^2+\frac{22434967}{7620480000}L\right)\nonumber \\
&&  + \frac{m^{10}}{M^{10}}\left(\frac{17}{300}\zeta_3
+ \frac{139}{111375}\zeta_2
- \frac{999168445440307}{14377502462400000}\right. \nonumber \\
&&  \left.  + \frac{128437}{149688000} L^2
+ \frac{1033765301}{691558560000} L \right)
+O\left(\frac{m^{12}}{M^{12}}\right).
\ee
These results where obtained in a
general covariant gauge, thus providing
additional checks at different steps of the calculation.

Let us discuss some general features of both series.
The leading term does not contain logarithms of the mass ratio
and represents therefore  a pure hard contribution. It is obtained from
the direct expansion of the graph in the small parameters of the problem:
the muon momentum $p$ and the muon mass $m$.
The integral is infrared finite and no soft subgraphs
appear in this order. It is represented by a one-scale integral with
the scale given by $M$.
In the next order of  $m^2/M^2$ soft subgraphs appear and,
consequently, logarithms of the mass ratio.
This is expected on general grounds:
A gauge invariant effective action 
proportional to $(F_{\mu\nu})^4/M^4$
--- the Euler-Heisenberg Lagrangian ---
can be constructed which is induced by heavy particles.
In all orders of the expansion the
maximal power of the logarithm is two, a consequence of the
singularities of the Feynman diagrams, which is reflected in the
structure of the expansion.

From Eqs.~(\ref{qedres1},\ref{scalqedres1}) it is apparent that the
result is sensitive to loop momenta of order $M$.
Otherwise the leading order would behave as $1/M^4$ and could be
easily obtained by using the local Euler-Heisenberg Lagrangian.

Numerically we get for the fermionic contribution
($x=m^2/M^2$)
\be
&&a_{\mu}(\gamma\gamma;{\rm QED}) = 0.6156 x \nonumber \\
&&+ (-0.1512 + 0.1666 \ln(x)-0.0497 \ln^2(x)) x^2 \nonumber \\
&&+ (-0.0453 + 0.0780 \ln(x)-0.0362 \ln^2(x)) x^3  \\
&&+ (-0.0194 + 0.0440 \ln(x)-0.0276 \ln^2(x)) x^4\nonumber \\
&&+ (-0.0099 + 0.0278 \ln(x)-0.0219 \ln^2(x)) x^5
+O(x^5).\nonumber
\ee
The series seems to converge even at the point $x=1$
where the sum of five terms
$a_{\mu}(m=M) = 0.616 - 0.151-0.045 -0.019-0.010\ldots$ leads to 0.39,
which has to be compared with the exact result
$a_{\mu}(m=M)|_{\rm exact}
= 0.3710\ldots$~\cite{Laporta:1991pa,Laporta:1992pa}.

In scalar QED the expansion has the following numerical form
\be
\label{scalNum}
&&a_{\mu}(\gamma\gamma;{\rm sQED}) = -0.0849 x \nonumber \\
&&+ (0.0220 - 0.0194 \ln(x) + 0.0052 \ln^2(x)) x^2 \nonumber \\
&&+ (0.0049 - 0.0067 \ln(x) + 0.0027 \ln^2(x)) x^3  \\
&&+ (0.0016 - 0.0029 \ln(x) + 0.0015 \ln^2(x)) x^4\nonumber \\
&&+ (0.0007 - 0.0015 \ln(x) + 0.0009 \ln^2(x)) x^5+O(x^5). \nonumber
\ee
At the point $m=M$  one finds
$a_{\mu}(\gamma\gamma;{\rm sQED}) =-0.0849+0.0220+0.0049
+0.0016+0.0007 = -0.0557$ with reasonably decreasing individual terms.
In this case the  exact result is unknown, it could, however
be calculated with the technique of ref.~\cite{Laporta:1992pa}.
The difference between the contributions of fermions
and scalars is apparent: for scalars
the result is negative and almost an order of magnitude smaller
in absolute value. The rate of convergence is quite similar in both cases.

In the present case of interest the actual value of
the expansion parameter is given by the ratio
$(m_\mu/m_\pi)^2=(106/139.6)^2=0.577$.
This leads to a rapidly converging series
$a_{\mu}(\gamma\gamma;{\rm sQED})=-0.0490\ (1-0.233 -0.037 -0.008-0.002)=- 0.0353$
with an estimated accuracy better than 0.5\%.
This value should be compared with the numerical evaluation of
ref.~\cite{Kinoshita:1984it}
where two methods were used with the results
$a_{\mu}(\gamma\gamma;{\rm sQED})|_{\rm num,1}=-0.0437(36)$
and
$a_{\mu}(\gamma\gamma;{\rm sQED})|_{\rm num,2}=-0.0383(20)$.
The slight disagreement between these two values is presumably caused
by, quoting the authors of ref.~\cite{Kinoshita:1984it}, ``a
delicateness of cancellation when separate contributions are put
together''.
The present approach is based on an analytical evaluation
that guarantees 
the absence of error accumulation. The accuracy of the result
is determined by the convergence rate of the series in
Eq.~(\ref{scalNum}) which is quite high for physical values of muon
and pion masses.

To summarize: A systematic expansion of the
light-by-light contribution to the anomalous magnetic moment of
the muon  has been presented which is valid for virtual fermions and
bosons with masses above or equal $m_\mu$.
The formulae are sufficiently accurate for the physical
applications of interest.

{\em Acknowledgements:}
The authors thank K.G.~Chetyr\-kin for fruitful discussions,
W.~Marciano for discussions and encouragement,
S.~Laporta for communications
and M.~Tentyukov for help with the computer system DIANA.
This work was supported in part by
DFG-Forschergruppe {\it ``Quantenfeldtheorie,
Computer\-algeb\-ra und Monte-Carlo-Simulation''}
(contract FOR 264/2-1),
by BMBF under grant No 05HT9VKB0, by Russian Fund for Basic
Research under contracts 01-02-16171 and 02-01-00601,
and by Volkswagen grant.

\end{document}